\documentclass[11pt]{article}               
\usepackage{osajnl}
\usepackage{amsmath,amssymb,graphicx}

\def\ds{\displaystyle}
\def\kperp{\mbox{$\rm\bf k_\perp$}}
\def\kperpp{\mbox{$\rm\bf k'_\perp$}}
\def\kperpd{\mbox{$\rm\bf k^2_\perp$}}
\def\kperpdp{\mbox{$\rm\bf k^{'2}_\perp$}}

\def\ezero{\mbox{${\rm\bf E_0}$}}
\def\dezero{\mbox{${\rm\bf e_0}$}}

\def\khatpw{\mbox{${\rm\bf\hat k_{pw}}$}}
\def\phatpw{\mbox{${\rm\bf\hat p_{pw}}$}}
\def\shatpw{\mbox{${\rm\bf\hat s_{pw}}$}}

\def\er{\mbox{ ${\rm\bf r}$}}
\def\rperp{\mbox{ ${\rm\bf r_\perp}$}}
\def\rperpd{\mbox{ ${\rm\bf r^2_\perp}$}}

\def\chat{\mbox{${\rm\bf\hat c}$ }}
\def\xhat{\mbox{${\rm\bf\hat x}$ }}
\def\yhat{\mbox{${\rm\bf\hat y}$ }}
\def\zhat{\mbox{${\rm\bf\hat z}$ }}
\def\xhatI{\mbox{${\rm\bf\hat x_I}$ }}
\def\yhatI{\mbox{${\rm\bf\hat y_I}$ }}
\def\zhatI{\mbox{${\rm\bf\hat z_I}$ }}

\begin{document}
\vspace*{-1.8cm}
\begin{flushright}
{\large\bf LAL 02-85}\\
\vspace*{0.1cm}
{\large September 2002}
\end{flushright}

\vspace{5mm}
 
\title{ Transmission and reflection of
 Gaussian beams   \\
by anisotropic parallel plates}

\author{Fabian Zomer}

\address{ Laboratoire de l'Acc\'el\'erateur Lin\'eaire, IN2P3-CNRS
 et Universit\'e de Paris-Sud,\\
 B.P. 34 - 91898 Orsay cedex, France.} 

\begin{abstract}
 Explicit and compact expressions 
 describing the reflection and the transmission of a Gaussian beam by
 anisotropic parallel plates
  are given. Multiple reflections inside the plate  
 are taken into account as well as arbitrary optical axis orientation and
 angle of incidence. 
\end{abstract}


\maketitle

\section{Introduction}

 Anisotropic parallel plates are extensively used in ellipsometry \cite{azzam}.
To precisely describe such experiments, it is necessary to take into
 account internal multiple reflections inside these plates 
\cite{azzam} and the Gaussian nature of 
laser beams \cite{bretagne}. However, to
 the author's knowledge, no general expression of a corresponding
 Mueller matrix can be found in the literature. 

In Ref. \citeonline{bretagne}, the problem is solved for Gaussian beams but in
 a particular case: uniaxial parallel plate tilted around
 the optical axis (itself located in the plane of
 incidence). Although very important results are provided, the formalism
 introduced by the authors cannot be generalized to an arbitrary geometrical 
 configuration, i.e. to a rotating tilted birefringent plate with an
  optical axis not necessarily in the plate interface. 
 Moreover, in this work, the calculations were
 carried out in the direct $(x,y,z)$ space. This feature has two major 
 implications: 1) effects related to the beam divergence 
 cannot be studied and more importantly, 2) the fact
 that a Jones matrix is only defined under a particular 
 approximation cannot be pointed out.

An adequate formalism to carry out the full calculations is indeed 
 the Fourier optics. Theoretical ground for the scalar and vector
 Fourier optics has been set up some time ago in a series of articles 
 \cite{fourier-scal,fourier-vec}.
 Thanks to this formalism and to the $4\times 4$ matrix
 method of Ref. \citeonline{yeh4x4},
 general and compact expressions describing the 
 transmission and reflection of a Gaussian beam by anisotropic parallel plates
 are provided. This is the topic of the present article.

 This paper is organized as follows.
 In the first section, useful features of the vector
 Fourier formalism are summarized. This formalism is then used
 in section \ref{sec-appl-vec} to derive a general expression for anisotropic
 parallel plates. Only the paraxial approximation is assumed at this stage. 
 A useful approximation, named
 here `scalar Fourier approximation', is then introduced in section
 \ref{sec-scalar}. This approximation, implicitly introduced in Ref.
 \citeonline{bretagne}, provides simpler formulas and the possibility to define
 an extended Mueller matrix for birefringent parallel plates.
 Numerical examples are finally given in section \ref{sec-uniaxe}. 
\newpage

\section{Vector Fourier optics in the paraxial approximation}\label{sec-vec}

All along this article we shall only be concerned with 
 lossless homogeneous anisotropic media and monochromatic Gaussian beam. 
In this section, we start by considering the propagation of a Gaussian beams
 in isotropic media. The main results obtained in 
Ref.  \citeonline{cross} are recalled
 together with additional information required for the following
 of the present paper.

\begin{figure}[h]
\begin{center}
\vspace{0.5cm}
\includegraphics[width=12cm]{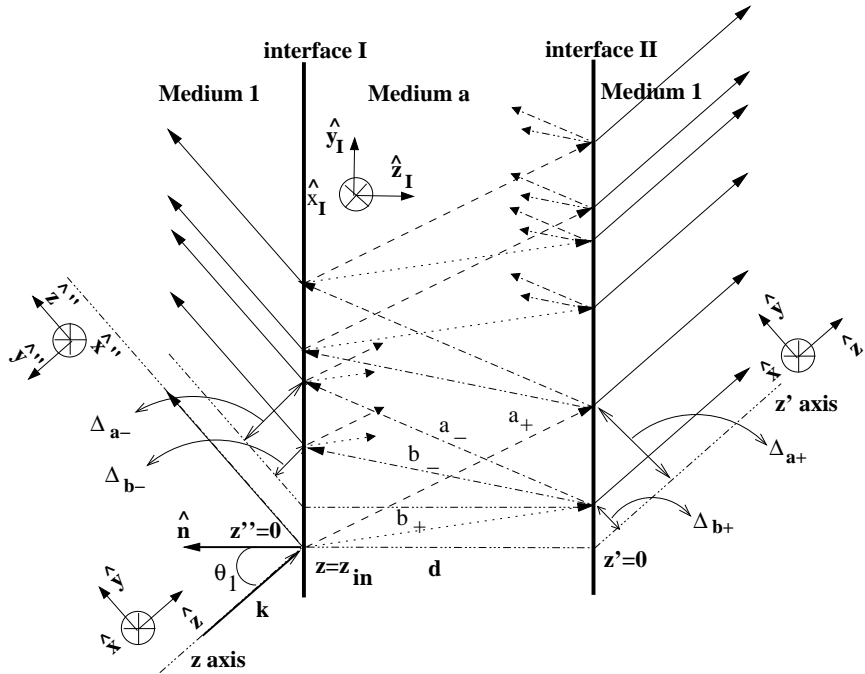}
\end{center}
\vspace{0.2cm}
\caption{Schematic view of the plane wave propagation in the anisotropic
 slab. For the sake of clarity, some of the inner reflected rays 
are represented by small arrows. The plane of
 incidence coincides with the plane $yz$. Symbols $a_\pm$ and
 $b_\pm$ correspond to the four possible propagation directions for
 the incident angle $\theta_1$. The four elementary
 transverse walk-offs, $\Delta_{a_\pm}$ and $\Delta_{b_\pm}$, are 
 indicated.
 The different vector basis used throughout this article are also shown:
 $\{\xhat,\yhat,\zhat\}$ for the incidesfigure4nt and transmitted beams,
 $\{\xhat'',\yhat'',\zhat''\}$ for
 the reflected beam, $\{\xhatI,\yhatI,\zhatI\}$ to perform calculations 
 of the birefringence effects. Attached to the two former bases are the 
 axis of propagation: $z$ axis for the incident beam, $z'$ axes for
 the transmitted beam and $z''$ axis for the reflected beam.}
    \label{zigzag}
\end{figure}

To describe the beam propagation,
 the direct system axis $(x,y,z)$ is chosen in such way that
 $\zhat={\rm\bf k}/k$ where  ${\rm\bf k}$ is the
 the Gaussian beam center's wave vector and $k=|{\rm\bf k}|$
 (see Fig. \ref{zigzag}). The
 origin $z=0$ is taken at the position where the beam size is minimum,
 i.e. at the beam waist position.
 The position vector will be
  written $\er=\rperp+z\zhat$ with $\rperp=x\xhat+y\yhat$ and where
 $\xhat$, $\yhat$ and $\zhat$ are unit vectors along $ox$, $oy$ and
 $oz$ axis respectively.

\newpage

In the paraxial approximation, an electromagnetic scalar 
field amplitude $\psi(\rperp,z=0)$ can be expanded according to
\begin{equation}\label{preums}
\psi(\rperp,0)={\cal F}[{\overline \psi}(\kperp)]=\frac{\ds 1}{\ds 2\pi}
\iint{\overline \psi}(\kperp)\exp(-i\kperp\cdot\er)d^2\kperp
\end{equation}
where the time dependence has been omitted and 
 where
 $\kperp=k_x \xhat+k_y\yhat$ and $|\kperp|\ll k$ so that
 $k_z\approx k(1-|\kperp|^2/(2k^2))$.
 In Eq. (\ref{preums}),
 ${\overline \psi}(\kperp)$ is the scalar field amplitude in the 
 $\kperp$ space. To satisfy the paraxial approximation,
 ${\overline \psi}(\kperp)$ must be such that the integral has
 appreciable values only for $|\kperp|\ll k$.

 At a given plane $z\ne 0$, the field is given by \cite{bacry}
\begin{equation}\label{evolu}
\psi(\rperp,z)=\exp\biggl(\frac{\ds -iz}{\ds 2k}
(P_x^2+P_y^2)\biggr)\psi(\rperp,0)
\end{equation}
where, as all along this article,
 the term $\exp(ikz)$ has been omitted and where 
\[
P_x=-i\frac{\partial}{\partial x},\,\,P_y=-i\frac{\partial}{\partial y}.
\]
Eq. (\ref{preums}) can then be written
\[
\psi(\rperp,z)={\cal F}[{\overline \psi}(\kperp,z)],
\]
where we define for convenience
\[
{\overline \psi}(\kperp,z)={\overline \psi}(\kperp)\exp\biggl
(\frac{\ds iz}{\ds 2k}\kperpd\biggr).
\]

For a paraxial Gaussian beam, it is easy to show that
\[
\psi(\rperp,0)=\sqrt{\frac{2}{\pi w_0^2}}\exp\biggl(\frac{\ds -\rperpd}
{\ds w_0^2}
\biggr)
\Longrightarrow \begin{cases}
\psi(\rperp,z)=\sqrt{\frac{2}{\pi w_0^2}}\frac{\ds -iz_R}{\ds q(z)}
\exp\biggl(\frac{\ds -ik\rperpd}{\ds 2q(z)}\biggr)\\
{\overline \psi}(\kperp,z)=\frac{w_0}{\sqrt{2\pi}}\exp\biggl(
\frac{\ds iq(z)\kperpd}{\ds 2k}\biggr)
\end{cases}
\] 
where $w_0$ is the beam waist, $q(z)=z+iz_R$ is the complex radius of
 curvature and $z_R=k w_0^2$ is the Rayleigh range. 

 Eq. (\ref{preums}) is the plane
 wave expansion of the vector Gaussian beam. Wave vectors of these
 plane waves are  defined by 
\begin{equation}\label{kperp}
{\rm\bf k_{pw}}=\kperp+k_z\zhat\approx\kperp+k(1-\frac{\ds |\kperp|^2}
{\ds 2k^2})\zhat
\end{equation} 
in the paraxial approximation.
 However, Eq. (\ref{preums}) does not take into account the vectorial nature
 of the electromagnetic field. To keep the orthogonality between the electric
field, magnetic field and  wave vector of each plane waves,
 one must introduce the six component vector \cite{cross} 
\[
F(\rperp,0)=
\begin{pmatrix}
E_x(\rperp,0)\\
E_y(\rperp,0)\\
E_z(\rperp,0)\\
B_x(\rperp,0)\\
B_y(\rperp,0)\\
B_z(\rperp,0)
\end{pmatrix}
\]
and use the following Fourier transformation \cite{cross} 
\begin{equation}\label{vector}
F(\rperp,0)=\frac{1}{2\pi}\iint
{\overline \psi}(\kperp)
 \exp(-i\kperp\cdot\rperp)
\exp\biggl(i\frac{\ds k_x}{\ds k}G_x+i\frac{\ds k_y}{\ds k}G_y\biggr)
\begin{pmatrix}
a_{1}\\
a_{2}\\
0\\
-a_{2}\\
a_{1}\\
0
\end{pmatrix}d^2\kperp
\end{equation}
where $G_x$ and $G_y$ are $6\times 6$ matrices which are derived from
 the Poincar\'e group algebra \cite{fourier-vec}.
 Expressions of these matrices
 can be found in Ref. \citeonline{cross}.
 It should be noticed that the orthogonality
 between the electric field, magnetic field and wave vector of the plane
 waves only holds in the paraxial approximation, i.e. up to the order
 ${\cal O}(|\kperp|^2/k^2)$.

Expression of $F(\rperp,z)$ is obtained similarly to Eq. (\ref{evolu})
\begin{equation}\label{vector2}
F(\rperp,z)=\frac{1}{2\pi}\iint
{\overline \psi}(\kperp,z)
 \exp(-i\kperp\cdot\rperp)
M_{6\times 6} 
\begin{pmatrix}
a_{1}\\
a_{2}\\
0\\
-a_{2}\\
a_{1}\\
0
\end{pmatrix}d^2\kperp,
\end{equation}
with
\[ 
M_{6\times 6}=
\left (\begin {array}{cccccc} 1+{\frac {{{ k_y^2}}
-{{ k_x^2}}}{8{k}^{2}}}&{\frac {{- k_y}{ k_x}}{4{k}^{2}}}
&{\frac {{ k_x}}{2k}}&{\frac {{ k_y}{ k_x}}{4{k}^{2}}}
&{\frac {{{ k_y^2}}-{{ k_x^2}}}{8{k}^{2}}}&{\frac 
{{- k_y}}{2k}}\\\noalign{\medskip}{\frac {-{ k_y}{ k_x}}{4{k
}^{2}}}&{1-\frac {{{ k_y^2}}-{{ k_x^2}}}{8{k}^{2}
}}&{\frac {{ k_y}}{2k}}&{\frac {{{ k_y^2}}-{{ k_x}}
^{2}}{8{k}^{2}}}&{\frac {-{ k_y}{ k_x}}{4{k}^{2}}}&{
\frac {{ k_x}}{2k}}\\\noalign{\medskip}{\frac {{- k_x}}{2k}}&
{\frac {{- k_y}}{2k}}&1&{\frac {{ k_y}}{2k}}&{\frac {{
- k_x}}{2k}}&0\\\noalign{\medskip}{\frac {{ -k_y}{ k_x}}{4{k
}^{2}}}&{\frac {-{{ k_y^2}}+{{ k_x^2}}}{8{k}^{2}}}&{
\frac {{ k_y}}{2k}}&1+{\frac {{{ k_y^2}}-{{ k_x
}}^{2}}{8{k}^{2}}}&{\frac {{- k_y}{ k_x}}{4{k}^{2}}}&{
\frac {{ k_x}}{2k}}\\\noalign{\medskip}{\frac {-{{ k_y^2}}+{
{ k_x^2}}}{8{k}^{2}}}&{\frac {{ k_y}{ k_x}}{4{k}^{2}}}&
{\frac {{ -k_x}}{2k}}&{\frac {-{ k_y}{ k_x}}{4{k}^{2}}}&
1-{\frac {{{ k_y^2}}-{{ k_x^2}}}{8{k}^{2}}}&
{\frac {{ k_y}}{2k}}\\\noalign{\medskip}{\frac {{- k_y}}{2k}}&
{\frac {{ k_x}}{2k}}&0&{\frac {{- k_x}}{2k}}&{\frac {
{- k_y}}{2k}}&1\end {array}\right ).
\]

\vspace{3mm}
\noindent
Focusing on the electric field,
 Eq. (\ref{vector2}) can be reduced to a $3\times 3$ matrix equation
\vspace{2mm}
\begin{eqnarray}\label{vector3}
\begin{pmatrix}
E_x(\er)\\
E_y(\er)\\
E_z(\er)
\end{pmatrix}
&=&{\cal F}[{\overline \psi}(\kperp,z) M_{3\times 3} \ezero]\\
&=&\frac{1}{2\pi}\iint
{\overline \psi}(\kperp,z)
 \exp(-i\kperp\cdot\rperp)
M_{3\times 3} 
\ezero
d^2\kperp\nonumber
\end{eqnarray}
\vspace{1mm}
with $\ezero^T=(a_1,a_2,0)$ and
\vspace{1mm}
\begin{equation}\label{m32}
M_{3\times 3} = 
\begin{pmatrix} 
1+{\frac {{{ k_y^2}}
-{{ k_x^2}}}{4{k}^{2}}}&
{\frac {{- k_y}{ k_x}}{2{k}^{2}}}&0\\
{\frac {{- k_y}{ k_x}}{2{k}^{2}}}&1-{\frac {{{ k_y^2}}
-{{ k_x^2}}}{4{k}^{2}}}&0\\
{\frac {{ -k_x}}{k}}&{\frac {{ -k_y}}{k}}&0\\
\end{pmatrix}
.
\end{equation}
Notices that the same expression holds for the magnetic field.
\newpage

One can verify that, for an electric vector polarized along $\xhat$ (i.e. 
$\ezero^T=(1,0,0)$), the integration of Eq. (\ref{vector3}) over $\kperp$
  leads to the results of
  Ref. \citeonline{cross}. The angular divergence
 of a Gaussian beam therefore generates crossed polarization effects.

\vspace{5mm}

\section{Application to anisotropic layer}\label{sec-appl-vec}

We shall now consider a Gaussian beam
 crossing a single anisotropic parallel
 plate of thickness $d$ surrounded by
 a dielectric medium of optical index $N_1$.
 The plate is located at $z=z_{in}$ and
 tilted with respect to ${\rm\bf k}$ (see Fig. \ref{zigzag}). The optical
 axis has an a priori arbitrary orientation.
 To compute the transmitted beam, the vector
 Fourier optics and the $4\times 4$ matrix formalism of Ref.
 \citeonline{yeh4x4}
 will be combined. In doing so, multiple reflections inside the parallel
 plate will be taken into account. 

 As indicated in Ref. \citeonline{simon}, polarization effects induced
 by the anisotropic plate
 can be computed by defining an operator acting on every
 plane waves constituting the Gaussian beam. This is further justified since
 we do only consider here homogeneous anisotropic media. Hence,
  writing ${\rm\bf E}_{t}(\er)$ 
 for the transmitted beam, one obtains \cite{simon,linear}
\begin{equation}\label{vector4}
{\rm\bf E}_{t}(\er')=\exp\biggl(\frac{-iz'}{2k}(P_x^2+P_y^2)\biggr)
{\cal F}\biggl[{\overline{ M}}_t 
\exp\biggl(\frac{-iz_{in}}{2k}(P_x^2+P_y^2)\biggr)
{\overline \psi}(\kperp,0) M_{3\times 3} 
\ezero
\biggr].
\end{equation}

 In Eq. (\ref{vector4}),  ${\overline{ M}}_t $ is a $3\times 3$
 matrix acting on the polarization
 state of each plane wave and $z'$ is an axis parallel to the $z$ axis
 with $z'=0$ at the exit of the plate (see Fig. \ref{zigzag}). This axis
 is used to describe the beam propagation after the plate. Inside the
 plate, the propagation of the plane waves is described by the matrix
 ${\overline{ M}}_t $. Let us mention that the position of the 
 $z'$ axis along the exit face of the plate can be arbitrarily chosen since
 it only introduces a global phase shift. 

However, in Eq. (\ref{vector4}) ${\overline{ M}}_t $ is determined in 
 the basis $\{\xhat,\yhat,\zhat\}$. For
 consistency with the general $4\times 4$ matrix
 method \cite{yeh4x4}, 
 ${\overline{ M}}_t $ must be determined in the plane wave
 polarization basis. The polarization vector basis is denoted
 by $\{\shatpw,\phatpw,\khatpw\}$ where 
\[
\shatpw=\frac{\khatpw\times{\rm\bf\hat n}}{|\khatpw\times{\rm\bf\hat n}|}
,\,\,\, \phatpw=\khatpw\times\shatpw 
\]  
correspond to
 the TE and TM waves respectively and where
 ${\rm\bf\hat n}$ is the unit vector normal 
 to the interface. The direction of the wave vector reads 
$\khatpw={\rm\bf k_{pw}}/k$ with ${\rm\bf k_{pw}}$ given by Eq. (\ref{kperp})
 and
 by convention, $\{\shatpw,\phatpw,\khatpw\}=\{\xhat,\yhat,\zhat\}$ when
 $\kperp=0$.

 The plane of incidence being related to the plane wave
 vector one gets
\begin{equation}\label{omega0}
{\overline{ M}}_t =\Omega  M_t \Omega^T 
\end{equation}
with
\begin{eqnarray}
\Omega  &=&
\begin{pmatrix}
\xhat\cdot\shatpw&\xhat\cdot\phatpw&\xhat\cdot\khatpw\\
\yhat\cdot\shatpw&\yhat\cdot\phatpw&\yhat\cdot\khatpw\\
\zhat\cdot\shatpw&\zhat\cdot\phatpw&\zhat\cdot\khatpw
\end{pmatrix}
,\label{omega}\\
M_t &=&\left(
\begin{array}{c c}
{\cal M}_t  &\begin{array}{c}0\\0\end{array}\\
\begin{array}{c c}0&0\end{array}&0\\
\end{array}
\right)\label{ucal}
\end{eqnarray}
and where ${\cal M}_t $ is the $2\times 2$ matrix describing the
 transmission of the plane waves in the $\{\shatpw,\phatpw\}$ basis.
 The matrix ${\cal M}_t $ is obtained by reducing the  $4\times 4$
 matrix method of Ref. \citeonline{yeh4x4} to a $2\times 2$ matrix
 algebra as described in the following section. 

\subsection{$2\times 2$ matrix algebra for plane wave propagation inside
 anisotropic parallel plates}
\vspace{5mm}
Because of linearity, the relations between incident, reflected and refracted
 plane waves amplitudes at an interface between two anisotropic media
 can be written in a $2\times 2$ matrix form.
 For the two interfaces I and II of the single layer of Fig. \ref{zigzag},
 we thus introduce the following $2\times 2$ matrices:
\begin{align}
\mbox{Reflection $1\rightarrow a$ at interface I }:
&\, {\cal R}_{1a_+}\label{coeff-1}\\
\mbox{Transmission $1\rightarrow a$ at interface I }:
&\, {\cal P}_{+}^{-1} {\cal T}_{1a_+}\\
\mbox{Reflection  $a\rightarrow 1$ at interface I }:
&\,{\cal P}^{-1}_{+}
{\cal R}_{a1_-}{\cal P}_{-}\\
\mbox{Transmission  $a\rightarrow 1$ at interface I }:
&\,{\cal T}_{a1_-}{\cal P}_{-}\\
\mbox{Reflection $a\rightarrow 1$ at interface II }:
&\,{\cal R}_{a1_+}\\
\mbox{Transmission $a\rightarrow 1$ at interface II }:
&\,  {\cal T}_{a1_+}\label{coeff-7}
\end{align}
where the $+$
 and $-$ subscripts refer to forward and backward beam propagations
 respectively. ${\cal R}$ and ${\cal T}$ matrices describe
 the reflection 
and transmission of the plane waves. ${\cal P}_{\pm}$ are the diagonal phase 
 shift matrices 
\[
{\cal P}_{\pm}=
\begin{pmatrix}
\exp(id{\rm\bf k}_{a_\pm}\cdot {\rm\bf\hat n})&0\\
0&\exp(id{\rm\bf k}_{b_\pm}\cdot {\rm\bf\hat n})\\
\end{pmatrix}
\]
where ${\rm\bf k}_{a_\pm}$ and  ${\rm\bf k}_{b_\pm}$
are the four possible wave vectors \cite{yeh4x4},
 inside the anisotropic medium,
 corresponding to the incident wave vector defined in Eq. (\ref{kperp}).
 In the above expression, 
 ${\rm\bf\hat n}$ is the unit vector normal to the interface. 

 All the matrices of Eqs. (\ref{coeff-1}-\ref{coeff-7})
 are different
 in the general case of a biaxial medium. They are determined by the 
 electromagnetic field continuity conditions at interfaces I and II 
 of Fig. \ref{zigzag} according to the method of Ref. \cite{yeh4x4}.

Writing ${\bf E}_{pw,i}$ and ${\bf E}_{pw,t}$ the electric field vectors
 for an incident plane wave at interface I and the transmitted plane wave
 at interface II, we obtain  
\[
{\bf E}_{pw,t}=
\biggl({\cal T}_{a1_+}{{\cal P}}_+^{-1}{\cal T}_{1a_+}+
{\cal T}_{a_+}{{\cal P}}_+^{-1}{\cal R}_{a1_-}
{{\cal P}}_-
{\cal R}_{a1_+}{{\cal P}}_+^{-1}
{\cal T}_{1a_+}+\cdots\biggr){\bf E}_{pw,i},
\] 
using Eqs. (\ref{coeff-1}-\ref{coeff-7}).
 This expression can be simplified utilising
 $\sum_{j=0}^{\infty}X^j=(1-X)^{-1}$:
\begin{equation}\label{Mt-plate}
{\cal M}_t =
{\cal T}_{a1_+}\biggl[1-{\cal P}^{-1}_+{\cal R}_{a1_-}
{\cal P}_-{\cal R}_{a1_+}
\biggr]^{-1} {\cal P}^{-1}{\cal T}_{1a_+},
\end{equation}
with ${\bf E}_{pw,t}={\cal M}_t {\bf E}_{pw,i}$. 

Let us mention that using the same procedure one can also
 determine the reflected electric vector
 at the interface I of  Fig. \ref{zigzag}.

  \vspace{8mm}
\subsection{Transmitted beam Intensity} 
\vspace{3mm}
 Turning back to the transmission of Gaussian beams,
  Eq. (\ref{vector4}) can be further simplified when one is interested
 by intensity measurements. For example, if
 an optical polarization component is located upward the anisotropic medium,
 the out-coming electric field reads
\[
{\bf E}_{out}(\er)
=\frac{w_0}{\sqrt{2\pi}}
{\cal F}\biggl[ \exp\biggl(
i\frac{(z_{in}+z')\kperpd}{k}\biggr)\exp\biggl(
\frac{-w_0^2\kperpd}{4}\biggr)
J
{\overline M}_t M_{3\times 3} 
\ezero
\biggr]
\] 
where 
\[
J=
\left(
\begin{array}{c c}
{\cal J} &\begin{array}{c}0\\0\end{array}\\
\begin{array}{c c}0&0\end{array}&0\\
\end{array}
\right)
\]
and ${\cal J}$ is the $2\times 2$
 Jones matrix corresponding to this component \cite{linear}. 
  The total intensity measured after this component is given by
\begin{eqnarray}
I_{out}&=&
\iint|{\rm\bf E}_{out}|^2d^2\rperp\nonumber\\
&=&
\frac{w_0^2}{(2\pi)^3}\idotsint
\exp\biggl(\frac{-w_0^2(\kperpd+\kperpdp)}{4}\biggr)
\exp\biggl(
i\frac{(z_{in}+z')(\kperpd+\kperpdp)}{k}\biggr)
\nonumber\\
& &\biggr[O_{3\times 3} \ezero
\biggl]
\cdot
\biggl[O^*_{3\times 3}(\kperpp)\ezero^*
\biggr]
\exp(i(\kperp-\kperpp)\cdot\rperp)
d^2\kperp d^2\kperpp d^2\rperp
\end{eqnarray}
\vspace{5mm}
with $O_{3\times 3} ={ J}{\overline M}_t M_{3\times 3} $
 and where the symbol * stands for the complex conjugate.

If the matrix ${ J}$ does not depend on the transverse
 spatial coordinates, the previous equation is simplified 
\begin{eqnarray}\label{master}
I_{out}=\frac{w_0^2}{2\pi}
\iint
\exp\biggl(\frac{-w_0^2\kperpd}{2}\biggr)
\biggr|O_{3\times 3} \ezero
\biggr|^2
d^2\kperp 
\end{eqnarray}
after integrations over $\rperp$ and 
 $\kperpp$ and using 
\[
\delta^2(\kperp-\kperpp)=(2\pi)^2\iint \exp(i(\kperp-\kperpp)\cdot\rperp)
d^2\rperp
\]
where $\delta^2(\kperp-\kperpp)$ is the Dirac distribution.

Eq. (\ref{master})
 shows explicitly that the total intensity depends only on the 
 waist and not on the beam size inside the anisotropic system. This 
 observation has been already made and 
 experimentally tested in Ref. \citeonline{bretagne}.

Up to now only the transmission has been
 considered. Equivalent expressions can be obtained for the reflection
 by simply replacing ${\cal M}_t$ by the extended Jones matrix
 describing the reflection.

Eq. (\ref{master}) was obtained under the paraxial approximation.
 This equation can be used as it is but, depending
on the required accuracy, one can perform further 
 simplifications: the plane wave approximation
 and the Scalar Fourier
 approximation. The latter is the topic of the following section.

\vspace{3mm} 

\section{Scalar Fourier Approximation}\label{sec-scalar}


 Although the vector Fourier optics is a useful formalism to 
 describe the Gaussian beam, the cross polarization effects are indeed
 very small \cite{cross} (though being observable but 
 essentially in extinction experiments\cite{simon}).
 In addition, the birefringence induced by the beam angular divergence 
 is also expected
 to be small, at least for realistic ellipsometry experiments.
\newpage
 To a good approximation, one
 can then assume that all plane waves constituting the Gaussian beam
 have the same wave vector ${\rm\bf k}$ (i.e. $\kperp=0$).  
 The polarization effects induced by a birefringent plate will then be only
 related to the direction of the Gaussian beam's center.

 This approximation thus amounts to account for the 
 Gaussian nature of the beam only in the calculation 
 of the interference pattern of the beam after the plate, as in Ref.
 \citeonline{bretagne}. After the anisotropic plate, the beam is made of 
   a sum of Gaussian beams transversally shifted
 by a distance (the transverse walk-off), induced by successive
  internal reflections (see Fig. \ref{zigzag}).
 Using Eqs. (\ref{coeff-1}-\ref{coeff-7}), the first transmitted
 beam can be written
\begin{align}
{\rm\bf E}_{1t}(\er')=&\biggl({\frac{2}{\pi\omega_0^2}}\biggr)^{1/2}
\frac{-iz_R}{q(z'+z_{in})}
\exp\biggl(\frac{-ikx^{'2}}{2q(z'+z_{in})}\biggr)
\biggl[\nonumber\\
 &{\cal T}_{a1_+}
\begin{pmatrix}
\exp\biggl(\frac{-ik(y'-\Delta_{a_+})^2}{2q(z'+z_{in})}\biggr)&0\\
0&\exp\biggl(\frac{-ik(y'-\Delta_{b_+})^2}{2q(z'+z_{in})}\biggr)\\
\end{pmatrix}
{\cal P}_+^{-1}
{\cal T}_{1a_+}\biggr]\dezero
\label{champs1}
\end{align}
with $\dezero^T=(a_1,a_2)$ and 
where $\Delta_{a_+}$ and $\Delta_{b_+}$ are the transverse walk-offs.
 In biaxial media there are four different elementary transverse walk-offs 
\[
{\Delta}_{a_\pm}=\frac{c}{\omega}\frac{N_1d\sin\theta_1\cos\theta_1}
{|{\overline{\rm\bf k}}_{a_\pm}\cdot {\rm\bf\hat n}|},\,\,
{\Delta}_{b_\pm}=\frac{c}{\omega}\frac{N_1d\sin\theta_1\cos{\theta_1}}
{|{\overline{\rm\bf k}}_{b_\pm}\cdot
{\rm\bf\hat n}|} ,
\]
 where ${\overline{\rm\bf k}}_{a_\pm}$ and  ${\overline{\rm\bf k}}_{b_\pm}$
 are the wave vectors inside the anisotropic plate
 corresponding to the incident wave vector ${\rm\bf k}$,
 i.e. the center of the Gaussian beam. We should specify that all the matrices 
 of Eqs. (\ref{coeff-1}-\ref{coeff-7}) are 
 also determined with respect to the direction of the 
 center of the Gaussian beam. 

When the transmission and reflection interface matrices are not diagonal
 it becomes difficult, if not impossible, to write a general formula for the 
 $n^{th}$ transmitted beam.
 However, using the following property of the Fourier transform
\begin{equation}\label{prop}
\psi(\rperp,z)={\cal F}[{\overline \psi}(\kperp,z)]\Rightarrow
 \psi(\rperp-\Delta\yhat,z)=
{\cal F}[{\overline \psi}(\kperp,z)\exp(i\Delta k_y)],
\end{equation}
and taking the inverse Fourier transform of Eq. (\ref{champs1}), one 
 obtains 
\begin{align}
{\cal F}^{-1}[{\rm\bf E}_{1t}(\er')]=&\biggl(\frac{w_0^2}{2\pi}\biggr)^{1/2}
\exp\biggl(i\frac{(z_{in}+z')\kperpd}{k}\biggr)\exp\biggl(
\frac{-w_0^2\kperpd}{4}\biggr)\biggl[\nonumber\\
 &{\cal T}_{a1_+}
\begin{pmatrix}
\exp(ik_y\Delta_{a_+})&0\\
0&\exp(ik_y\Delta_{b_+})\\
\end{pmatrix}
{\cal P}_+^{-1}
{\cal T}_{1a_+}\biggr]\dezero.
\nonumber
\end{align}
It is therefore possible to sum up all the transmitted beams in the $\kperp$
 space by introducing a new set of $2\times 2$ matrices:
\begin{equation}\label{wmat}
{\cal W}_\pm=
\begin{pmatrix}
\exp(i{\Delta}_{a_\pm}k_y)&0\\
0&\exp(i{\Delta}_{b_\pm}k_y)
\end{pmatrix}
\end{equation}
describing the transverse walk-off of the Gaussian beam's center
 (see Fig. \ref{zigzag}). \\
\noindent
Following the method introduced in the 
 previous section, one then obtains 
\begin{eqnarray}
{\rm\bf E}_{t}(\er')
&=&\biggl(\frac{w_0^2}{2\pi}\biggr)^{1/2}{\cal F}\biggl[\exp\biggl(
i\frac{(z_{in}+z')\kperpd}{k}\biggr)
\exp\biggl(
\frac{-w_0^2\kperpd}{4}\biggr)
\biggl({\cal T}_{a1_+}\widetilde{{\cal P}}_+^{-1}{\cal T}_{1a_+}+
\nonumber\\
& &{\cal T}_{a_+}\widetilde{{\cal P}}_+^{-1}{\cal R}_{a1_-}
\widetilde{{\cal P}}_-
{\cal R}_{a1_+}\widetilde{{\cal P}}_+^{-1}
{\cal T}_{1a_+}+\cdots\biggr)
\dezero
\biggr]
\nonumber\\
&=&\biggl(\frac{w_0^2}{2\pi}\biggr)^{1/2}{\cal F}\biggl[\exp\biggl(
i\frac{(z_{in}+z')k_y^2}{k}\biggr)
\exp\biggl(
\frac{-w_0^2k_y^2}{4}\biggr)
\widetilde{{\cal M}}_t 
\dezero
\biggr],
\label{transm}
\end{eqnarray}
with
\begin{eqnarray}
\widetilde{{\cal P}}_+^{-1}&=&{\cal W}_+{\cal P}_+^{-1}\label{ptilde1}\\
\widetilde{{\cal P}}_-&=&{\cal W}_-{\cal P}_-\label{ptilde2}\\
\widetilde{{\cal M}}_t &=&
{\cal T}_{a1_+}\biggl[1-\widetilde{{\cal P}}^{-1}_+{\cal R}_{a1_-}
\widetilde{{\cal P}}_-{\cal R}_{a1_+}
\biggr]^{-1} \widetilde{{\cal P}}^{-1}_+{\cal T}_{1a_+}.\label{Mt-plate2}
\end{eqnarray}


In the scalar Fourier approximation 
 the usual Stokes vectors \cite{huard} can be defined
\[
S_{in}=
\begin{pmatrix}
|a_1|^2+|a_2|^2\\
|a_1|^2-|a_2|^2\\
a_1a_2^*+a_1^*a_2\\
i(a_1a_2^*-a_1^*a_2)
\end{pmatrix},\,\,
S_{out}=
\begin{pmatrix}
\iint(|E_{tx}|^2+|E_{ty}|^2)d^2\rperp\\
\iint(|E_{tx}|^2-|E_{ty}|^2)d^2\rperp\\
\iint(E_{tx}E_{ty}^*+E_{tx}^*E_{ty})d^2\rperp\\
i\iint(E_{tx}E_{ty}^*-E_{tx}^*E_{ty})d^2\rperp
\end{pmatrix},\,\,
\]
where $S_{in}$ and $S_{out}$ are related to the incident beam and
transmitted beam respectively. From Eq. (\ref{transm}) and 
 following the calculation steps of section \ref{sec-appl-vec}
one can then determine the extended Mueller matrix of the plate $M_S$
 such $S_{out}=M_SS_{in}$:
\vspace{6mm}

\begin{equation}\label{mueller}
M_S=
\begin{pmatrix}
(\rho^2_{11}+\rho^2_{12})/2+&
(\rho^2_{11}+\rho^2_{21})/2-&
\rho_{11,12}\cos\phi_{11,12}+&
\rho_{11,12}\sin\phi_{11,12}+\\
(\rho^2_{21}+\rho^2_{22})/2&
(\rho^2_{12}+\rho^2_{22})/2
 &\rho_{21,22}\cos\phi_{21,22}&\rho_{21,22}\sin\phi_{21,22}\\
& & & & \\
(\rho^2_{11}+\rho^2_{12})/2-&
(\rho^2_{11}+\rho^2_{22})/2-&
\rho_{11,12}\cos\phi_{11,12}-&
\rho_{11,12}\sin\phi_{11,12}-\\
(\rho^2_{21}+\rho^2_{22})/2
&(\rho^2_{21}+\rho^2_{12})/2 &
 \rho_{21,22}\cos\phi_{21,22}&\rho_{21,22}\sin\phi_{21,22}\\
& & & & \\
\rho_{11,21}\cos\phi_{11,21}+&
\rho_{11,21}\cos\phi_{11,21}-&
\rho_{11,22}\cos\phi_{11,22}+&
\rho_{11,22}\sin\phi_{11,22}-\\
\rho_{12,22}\cos\phi_{12,22}&\rho_{12,22}\cos\phi_{12,22}&
\rho_{12,21}\cos\phi_{12,21}&\rho_{12,21}\sin\phi_{12,21}\\
& & & & \\
-\rho_{11,21}\sin\phi_{11,21}-&
-\rho_{11,21}\sin\phi_{11,21}+&
-\rho_{11,22}\sin\phi_{11,22}-&
\rho_{11,22}\cos\phi_{11,22}-\\
\rho_{12,22}\sin\phi_{12,22}&\rho_{12,22}\sin\phi_{12,22}&
\rho_{12,21}\sin\phi_{12,21}&\rho_{12,21}\cos\phi_{12,21}\\
\end{pmatrix}
\end{equation}
with
\begin{eqnarray}
\rho^2_{ij}&=&\frac{w_0}{\sqrt{2\pi}}\int
\exp\biggl(\frac{-w_0^2k_y^2}{2}\biggr)
|m_{ij}|^2dk_y,\label{int1}\\
\rho_{ij,kl}&=&\frac{w_0}{\sqrt{2\pi}}\biggl|\int
\exp\biggl(\frac{-w_0^2k_y^2}{2}\biggr)
 m_{ij}m^*_{kl}dk_y\biggr|\nonumber\\
\cos\phi_{ij,kl}&=&\frac{1}{2\rho_{ij,kl}}\int
\exp\biggl(\frac{-w_0^2k_y}{2}\biggr)
[m_{ij}m^*_{kl}+m^*_{ij}m_{kl}]dk_y
\nonumber\\
\sin\phi_{ij,kl}&=&\frac{1}{2i \rho_{ij,kl}}\int
\exp\biggl(\frac{-w_0^2k_y}{2}\biggr)
[m_{ij}m^*_{kl}-m^*_{ij}m_{kl}]dk_y
\label{int4}
\end{eqnarray}
and $i,j,k,l=1,2$. The matrix elements $m_{ij}$ are defined by
\[
{\cal J}\widetilde{{\cal M}}_t=
\begin{pmatrix}
m_{11}&m_{12}\\
m_{21}&m_{22}
\end{pmatrix}.
\]
One feature related to the integral form of Eqs. (\ref{int1}-\ref{int4})
 is that
 $\rho_{ij,kl}\ne \rho_{ij}\rho_{kl}$. This means that 
 one cannot define a Jones matrix (integrated over $\kperp$).
\vspace{8mm}
\section{Example: uniaxial parallel plate}\label{sec-uniaxe}
\vspace{3mm}
 To illustrate our model we shall now consider the very
 important example of a quarter wave Quartz plate
 (QWP) \cite{handbook}.
The equivalence between our approach and the results of
 Ref. \citeonline{bretagne}
 is formally proved in Appendix I and details concerning the evaluation 
 of Eq. (\ref{master}) are given in Appendix II.

To study the accuracy of the plane wave and
 scalar Fourier approximations we shall adopt one of the examples of 
 Ref. \citeonline{bretagne}: a Gaussian laser beam of wavelength
 $\lambda=0.6328\,{\rm \mu m}$ and waist $w_0=100\,{\rm \mu m}$. For
 this wavelength,
 the Quartz ordinary and extraordinary optical indices are
 $n_0=1.542637$ and $n_e=1.551646$.

 Since interference effects are sensitive to the plate thickness, we choose
 to compare two realistic components:
 a first order QWP ($d=87.6010 \,{\rm \mu m}$) and a tenth order QWP
 ($d=719.9686 \,{\rm \mu m}$).
 Finally, two different polar orientations of the 
 optical axis (see Appendix II) are chosen $\theta_c=\pi/2$ (= optical
 axis in the plane of interface) and $\theta_c=\pi/4$. The remaining 
 geometric degrees of freedom are the optical axis azimuth $\phi_c$ and
 the incidence angle of the center of the Gaussian beam $\theta_1$.

 The incident
 beam is assumed to be polarized along $ox$ so 
 that $\ezero= \xhat$ (or $S_{in}=(1,1,0,0)$).
 As in Fig. \ref{zigzag}, the beam crosses a Quartz
 plate and we shall first consider that an intensity measurement is
 performed after
 a perfect linear polarizer aligned along $ox$. We shall denote
 $I_{||,Gauss}$ and $I_{||,pw}$ the corresponding
 intensities computed according to
 Eq. (\ref{mueller}) (scalar Fourier approximation) and in the plane
 wave approximation.
 We checked numerically
 that results obtained in the scalar Fourier approximation
 agree with the general expression of \ref{master} (to observe noticeable
 differences one must consider beam waits as small as $10 \,{\rm \mu m}$
 which are outside the scope of this article).

The relative numerical precision of the results presented below
 has been estimated to
 be of the order of $10^{-6}$. This number
 was determined by checking the energy
 conservation and by looking at the difference between the plane wave and
 the scalar Fourier approximation at normal incidence (they are similar
 by construction).

\begin{figure}[h]
\begin{center}
\vspace{-5mm}
\includegraphics[width=12cm]{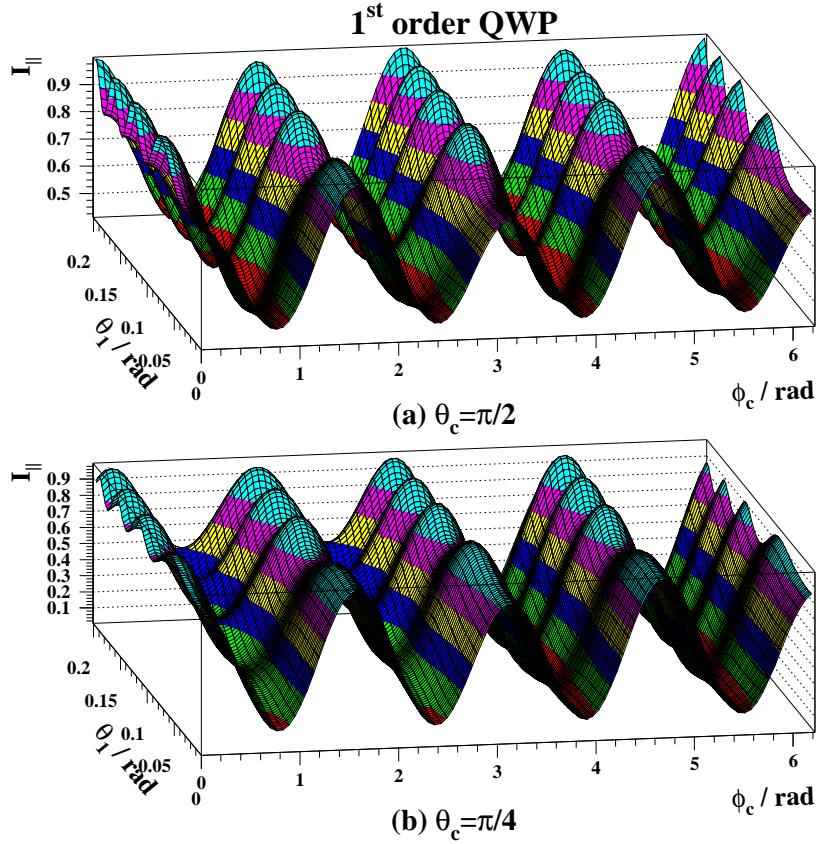}
\caption{Intensity of a Gaussian beam measured after a first
 order QWP and a perfect linear polarizer: (a) for an optical axis in the 
 plane of interface, (b) for an optical axis inclined by $\pi/4$ with respect 
 to the plane of interface. The calculations are performed using the 
 scalar Fourier approximation and are shown as function of the angle 
 of incidence $\theta_1$ and the azimuth angle of the optical axis $\phi_c$.   }
\label{intensite-2d-1}
  \end{center}
\end{figure}

\begin{figure}[hbt]
  \begin{center}
\includegraphics[width=11cm]{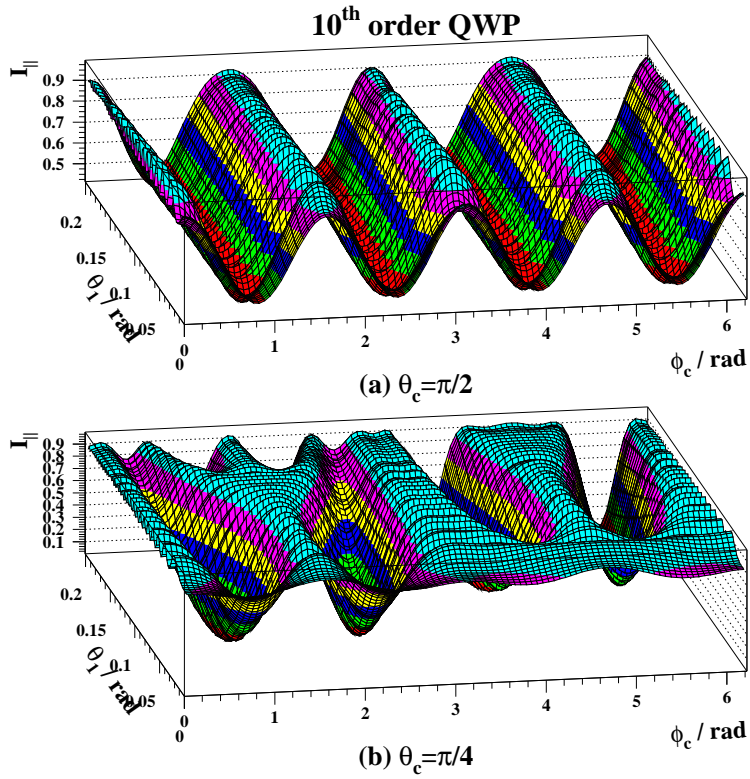}
\caption{Same as Fig.2 for a tenth order QWP.}
 \label{intensite-2d-10}
 \end{center}
\end{figure}

 $I_{||,Gauss}$ is shown as function of $\phi_c$ and $\theta_1$
in Fig. \ref{intensite-2d-1}(a-b) and \ref{intensite-2d-10}(a-b)
for the two plate thickness and the two orientations of the optical axis.\\

\noindent

As expected \cite{bretagne}, the interference pattern is denser for the tenth
 order plate (Fig. \ref{intensite-2d-10}(a) and (b)) and the intensity
 is not $\pi$ symmetric in $\phi_c$ when $\theta_c<\pi/2$ and $\theta_1\ne 0$ 
(Fig. \ref{intensite-2d-1}(b) and \ref{intensite-2d-10}(a)).

\newpage

\begin{figure}[h]
\begin{center}

\includegraphics[width=12cm]{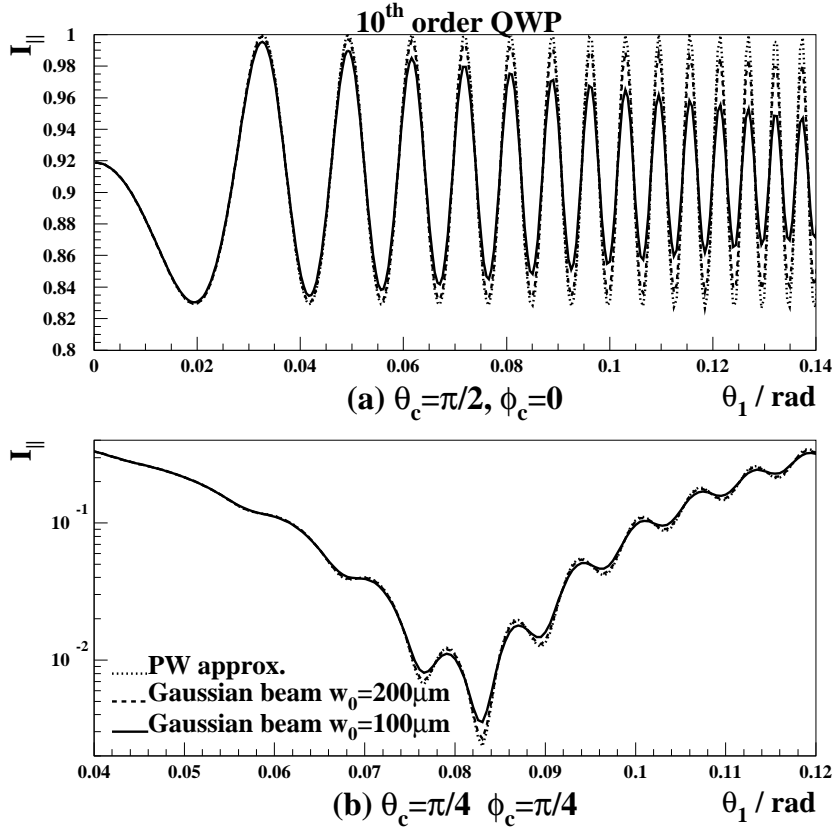}

\caption{Intensity of a Gaussian beam measured after a tenth
 order QWP and a perfect linear polarizer as function of the angle 
 of incidence $\theta_1$. Orientations of the QWP optical axis are fixed to:
 (a) $\theta_c=\pi/2$ and $\phi_c=0$, (b)
 $\theta_c=\pi/4$ and $\phi_c=\pi/4$. Full lines and dashed lines
 represent the calculations
 performed using the scalar Fourier approximations for $w_0=100\,{\rm m\mu}$
 and $w_0=200\,{\rm m\mu}$ respectively. Dotted lines show the calculations
 performed using the plane wave approximation.}

    \label{intensite-1d}
  \end{center}
\end{figure}

Fixing $(\theta_c=\pi/2,\phi_c=0)$ and $(\theta_c=\pi/4,\phi_c=\phi/4)$, 
 $I_{||,Gauss}$ and$I_{||,pw}$ are plotted as function of
 $\theta_1$ in Fig. \ref{intensite-1d} (a), (b) respectively
 for the tenth order QWP. In these figures, $I_{||,Gauss}$ is also computed
 for two beam waists $w_0=100\,{\rm \mu m}$ and $200 \,{\rm \mu m}$.
 Sizable differences appear which increase with the incident angle but
 decrease when the beam waist increases. 
 
\vspace{5mm}

\begin{figure}[h]
  \begin{center}
\includegraphics[width=12cm]{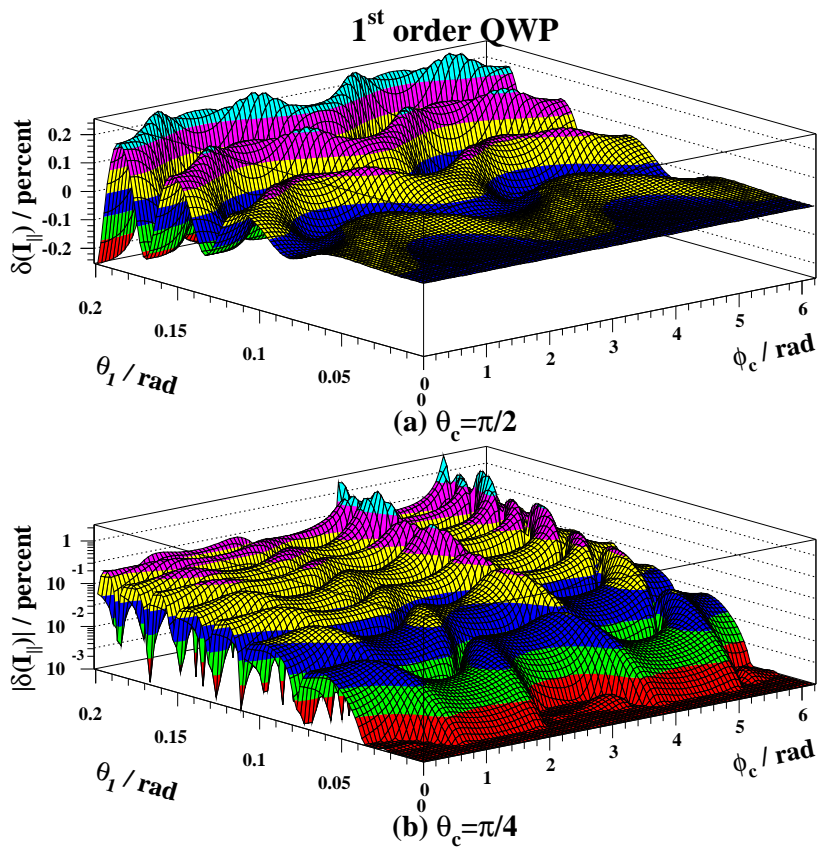}
\caption{ Relative difference between the quantity of
Fig.2 using the 
 plane wave approximation and the scalar Fourier approximation.}

    \label{approx-1}
  \end{center}
\end{figure}

\begin{figure}
  \begin{center}
\includegraphics[width=12cm]{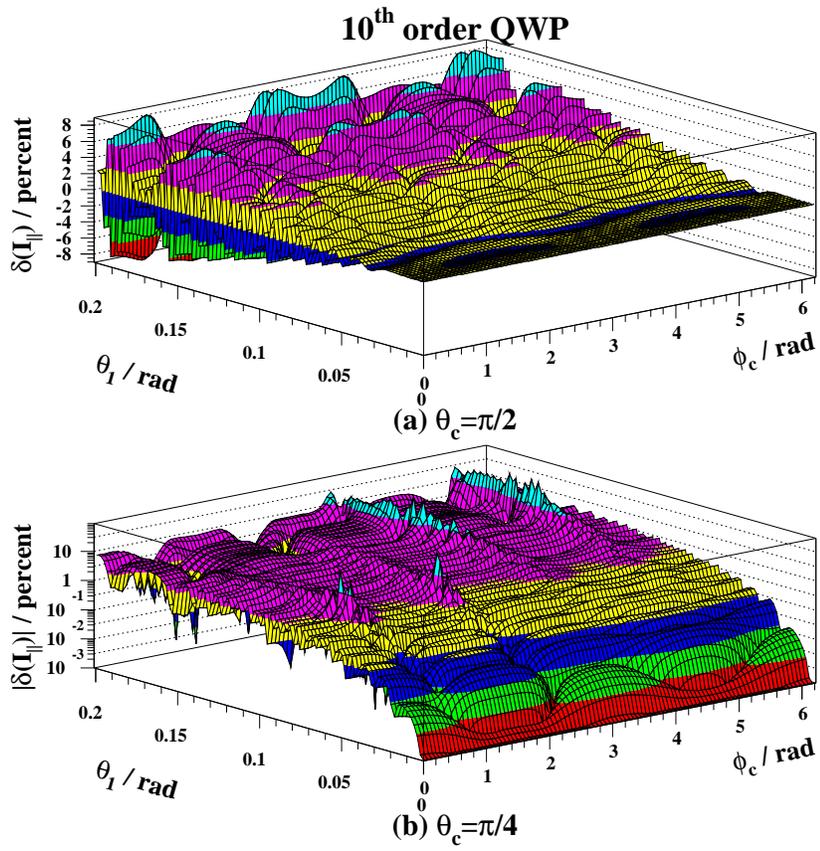}
\caption{Same a Fig.5 but the for the quantity of 
Fig.3.}
\label{approx-10}
  \end{center}
\end{figure}

To quantify the differences the following ratio is computed 

\[
\delta(I_{||})=\frac{I_{||,Gauss}-I_{||,pw}}{I_{||,Gauss}}
\] 

and plotted as function of $\phi_c$ and $\theta_1$ in Fig. 
Fig. \ref{approx-1}(a-b) and \ref{approx-10}(a-b)
for the two plate thicknesses and the two orientations of the optical axis.
One can sees that $\delta(I_{||})$ increases with the plate thickness
 and the angle
 of incidence. At $\theta_1\approx 0.2$~rad and $\theta_c=\pi/2$,
 large differences of
 the order of 10$\%$ appear for 
 the tenth QWP. Variations of  $\delta(I_{||})$
 with $\phi_c$ are also sizable. Especially for $\theta_c<\pi/4$ where 
 interference amplitudes are badly reproduced by the plane wave approximation
 for small values of $I_{||}$.\\
\newpage




\begin{figure}[h]
  \begin{center}
\includegraphics[width=10.1cm]{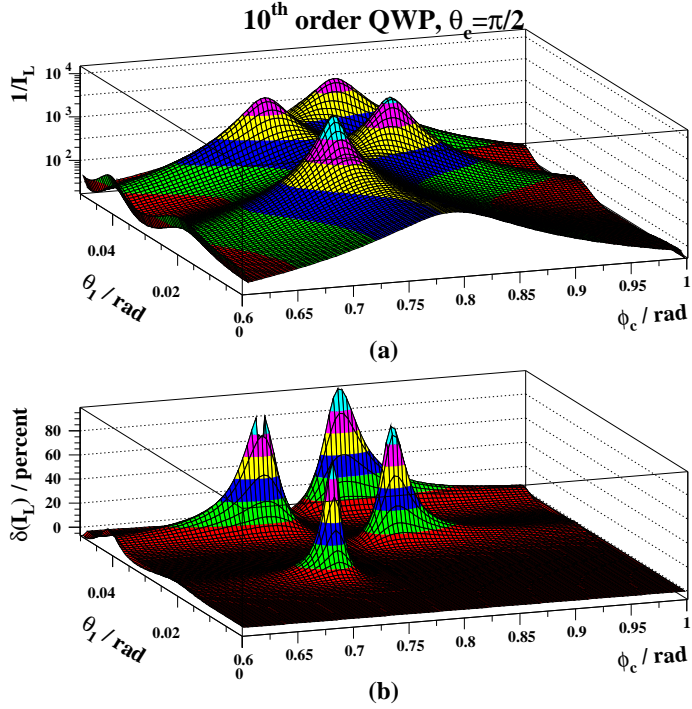}
\caption{(a) Inverse of the intensity measured after a tenth order 
 QWP and a perfect circular left polarizer. Calculations are performed
 using the scalar Fourier approximation and are shown as function of 
the incident angle $\theta_1$ and the azimuth angle of the optical axis
 $\phi_c$. (b) Relative difference between the quantity shown in the top
 plot calculated using the scalar Fourier approximation and the plane
 wave approximation. For these figures, the optical axis is taken in 
the plane of interface ($\theta_c=\pi/2$).}
    \label{sigmar}
  \end{center}
\end{figure}

\vspace{5mm}
Another interesting quantity is the degree of circular polarisation
 when the beam passes through a perfect circular polarizer, instead of the 
  linear polarizer of the above example.
 If the polarizer is circular left, using Eq. (\ref{mueller})
 and the standard Mueller matrices for perfect polarizers \cite{huard}
 the beam intensity reads 
\vspace{5mm}
\[
I_{L}=\frac{\rho^2_{11}+\rho^2_{21}}{2}-\rho_{11,21}\sin\phi_{11,21}
\]
for $S_{in}=(1,1,0,0)$ and $\theta_c=\pi/2$.
 When $\phi_c\approx\pi/4$, $I_L$ is small and
 can be minimized in the $(\theta_1,\phi_c)$ space 
 \cite{bretagne}. We present here $I_L^{-1}$ as function of $\theta_1$ and
 $\phi_c$ for the tenth order QWP in Fig. \ref{sigmar}(a).\\
\noindent
 Results of
 Ref. \citeonline{bretagne}
 are recovered although rotating the plate and rotating 
 the polarization, as done in this reference, are not strictly equivalent
 \cite{zhu,yeh2x2,yeh2x2-ax}.   
 In Fig. \ref{sigmar}(b) the relative difference between plane wave and
 Gaussian beam intensities
\[
\delta(I_L)=\frac{I_{L,Gauss}-I_{L,pw}}{I_{L,Gauss}}
\]  
is shown. Large differences corresponding to small 
 values of $I_L$ are observed. This demonstrate the necessity to account
 for the Gaussian nature of the beam in such particular, but important, cases. 
 It is to mention that the scalar Fourier approximation and the general 
 paraxial calculations are here also in perfect agreement.

\newpage
Finally the ratio $\rho_{11}\rho_{21}/\rho_{11,21}$ is computed. It increases
 with $\theta_1$ and the plate thickness but, even for the 
 tenth order QWP, it does not exceed 0.2~$\%$ for any values of $\phi_c$ and
 $\theta_1=0.2$~rad.
 Assuming $\rho_{11}\rho_{21}\approx\rho_{11,21}$,
 a Jones matrix can thus be defined
 with an accuracy of the order of a few per mill.

From this study one may conclude that:
\begin{itemize}
\item
the scalar Fourier approximation is very accurate with respect to 
 the paraxial approximation when the beam waist is not too small;
\item
the plane wave approximation mainly holds for thin Quartz plates, small
 incident angles, large beam waists
 and optical axes nearly in the interface plane. If these conditions
 are not fulfilled, an account for the Gaussian nature of the beam is
 necessary.
\item
It is crucial to account for the Gaussian nature of the beam when high
 performances of QWP are foreseen. 
\end{itemize}
More generally, accuracies of the various approximations strongly depend
 on the medium birefringence and geometrical configurations.

\section{conclusion}

General expressions describing the transmission and reflection of a Gaussian
 beam by anisotropic parallel plates have been obtained. The vector
 Fourier optics formalism of Refs. \citeonline{fourier-vec,cross}
 is used as general frame work.
 Multiple reflections inside the anisotropic medium
 are taken into account using a $2\times 2$ 
 matrix algebra derived from the general
 $4\times 4$ matrix formalism of Ref.\citeonline{yeh4x4}. The only assumption 
 supplied for these calculations is the paraxial approximation. 
 
 To simplify the calculations, a useful approximation introduced in 
 Ref. \citeonline{bretagne} is then considered. This approximation consist in 
 taking into account the Gaussian nature of the beam only in the interference
 pattern of the transmitted (or reflected) beam. Birefringence effects 
 induced by the beam angular divergence are then neglected. Accuracy
 of this approximation was checked for the particular case of uniaxial 
 Quartz parallel plates. For not too small beam waists,
 no noticeable differences were observed compared to the general 
 expression.

Precision of the plane wave approximation has also been checked using the
 example of Quartz plates. Here noticeable differences were observed. 
  These discrepancies do not trivially depend on the geometrical parameters. 
 Roughly, we concluded that
 they increase with the plate thickness and the angle of incidence. In the
 case of ellipsometry where a high purity circular polarization is foreseen,
 it was shown that an account for the Gaussian nature of the beam is
 necessary.

 As a general remark, accuracies of the various approximations presented in
 this article decisively depend
 on the birefringence of the medium, laser wavelength, geometrical
 configuration and type of energy measurements.
 They must then be checked case by case.  

 Interference effects, observed in the variations of the transmitted beam 
 intensity as function of the angle of incidence and optical axis azimuth, 
 suggest that a very precise calibration of a birefringent plate can be 
 performed without any other optical components. 

\section*{Acknowledgement}
I would like to thank V. Soskov for very stimulating discussions
 and suggestions, F. Marechal,
 C. Pascaud and N. Pavloff for careful reading and 
 discussions. 

\section{Appendix I}
In Ref. \citeonline{bretagne}, the transmission of a Gaussian beam though
 a tilted Quartz plate has been determined. The configuration
 is restricted to a tilt axis perpendicular or parallel to the optical axis, 
itself located in the interface. In this
 appendix we show the equivalence between the results
 of Ref. \citeonline{bretagne} and the formalism of section \ref{sec-scalar}. 

The elements of the transmission matrix are written
\begin{align}
m_{oo}&=\frac{t_{1{o}}t_{{o}1}\exp(-i(k_y\Delta_{o}+\delta_{o}))}{1-r_o^2
\exp(-2i(k_y\Delta_{o}+\delta_{o}))}\nonumber\\
m_{ee}&=\frac{t_{1{e}}t_{{e}1}\exp(-i(k_y\Delta_{e}+\delta_{e}))}{1-r_e^2
\exp(-2i(k_y\Delta_{e}+\delta_{e}))}\nonumber\\
m_{oe}&=m_{eo}=0\nonumber
\end{align}
where $e$ and $o$ refer to the extraordinary and ordinary wave respectively.
 The non zero extended Mueller matrix elements are given by:
\begin{align}
|\rho_{oo}|^2&=\sqrt{\frac{w_0^2k^2}{2\pi}}
\int |m_{oo}|^2\exp\biggl(\frac{-w_0^2k_y^2}{2}\biggr)dk_y\label{m11}\\
|\rho_{ee}|^2&=\sqrt{\frac{w_0^2k^2}{2\pi}}
\int|m_{ee}|^2\exp\biggl(\frac{-w_0^2k_y^2}{2}\biggr)dk_y\label{m22}\\
\rho_{oo;ee}\sin\phi_{oo;ee}&=
 \frac{1}{2i}\int [m_{oo}m_{ee}^*-m_{oo}^*m_{ee}] 
\exp\biggl( \frac{-w_0^2k_y^2}{2}\biggr)dk_y
\nonumber\\
&=\sqrt{\frac{w_0^2k^2}{2\pi}}
\int[\Re(m_{ee})\Im(m_{oo})-
\Re(m_{oo})\Im(m_{ee})]\exp\biggl( \frac{-w_0^2k_y^2}{2}\biggr)dk_y
\label{m12}\\
\rho_{oo;ee}\cos\phi_{oo;ee}&=
 \frac{1}{2}\int [m_{oo}m_{ee}^*+m_{oo}^*m_{ee}] 
 \exp\biggl(\frac{-w_0^2k_y^2}{2}\biggr)dk_y
\nonumber\\
&=\sqrt{\frac{w_0^2k^2}{2\pi}}
\int[\Re(m_{ee})\Re(m_{oo})+
\Im(m_{oo})\Im(m_{ee})] \exp\biggl(\frac{-w_0^2k_y^2}{2}\biggr)dk_y
\nonumber
\end{align}
with $m_{ij}=\Re(m_{ij})+i\Im(m_{ij})$ and 
 where $r_e$, $r_o$, $t_{e1}$, $t_{o1}$, $t_{1e}$, $t_{1o}$
  are the Fresnel coefficients for the particular geometric
 configuration studied here
 (expressions can be found in Ref. \citeonline{zander}).

In Ref. \citeonline{bretagne} a Jones matrix
\begin{equation}\label{rennesm}
\begin{pmatrix}
\sqrt{2a_1}&0\\
0&\sqrt{2a_2}\exp\biggl(
-i{\rm arcsin}\frac{a_3}{\sqrt{a_1a_2}}
\biggr)\\
\end{pmatrix}
\end{equation}
is defined and expressions for $a_1$, $a_2$ and $a_3$ are provided
 (Eqs. (A16a), (A16b) and (A16c) in Appendix I of Ref. \citeonline{bretagne}).

To demonstrate
the equivalence between our approach and the one of Ref. \citeonline{bretagne} 
 we must show that 
 Eqs. (\ref{m11}), (\ref{m22}) and (\ref{m12}) are integral forms of
 $2a_1$, $2a_2$ and $2a_3$ respectively. 
In order to prove it one just has to expand the integral kernels and then
 perform the integration over
 $k_y$. Since the
 scalar Fourier approximation was implicitly assumed in Ref.
 \citeonline{bretagne}, 
 one finds:
\begin{align}
\frac{|\rho_{oo}|^2}{2}=&\frac{t^2_x}{2}\biggl(1+
r_x^2\biggl[1+2\cos(4\varphi_x)\exp\biggl(\frac{-2\Delta^2}{w_0^2}\biggr)
\biggr]+\nonumber\\
&2r_x^3\biggl[\cos(2\varphi_x)
\exp\biggl(\frac{-\Delta^2}{2w_0^2}\biggr)+
\cos(6\varphi_x)\exp\biggl(\frac{-9\Delta^2}{2w_0^2}\biggr)\biggr]\nonumber\\
&+r_x^4\biggl[1+2\cos(4\varphi_x)\exp\biggl( \frac{-2\Delta^2}{w_0^2}\biggr)
+2\cos(8\varphi_x)\exp\biggl( \frac{-8\Delta^2}{w_0^2}\biggr)
\biggr]+\cdots\biggl)\label{A16a}
\end{align}
where notations of Ref. \citeonline{bretagne} $r_x=r_o^2$, $t_x=t_{1o}t_{o1}$,
 $\varphi_x=\delta_e$, $\varphi_x=\delta_o$ and
 $\Delta\approx 2\Delta_o\approx 2\Delta_e$
  have been adopted.
  Eq. (\ref{A16a}) can be further written as a series, and
 this leads to Eq. (A16a) of Ref. \citeonline{bretagne}.

 The series expansion of $|\rho_{ee}|^2/2$ is obviously obtained
 by replacing $r_x$ and $t_x$
 by $r_y=r_e^2$ and $t_y=t_{1e}t_{e1}$ respectively in Eq. (\ref{A16a}). This 
 leads to Eq. (A16b) of Ref. \citeonline{bretagne}.

 In the same way one finally obtains
\begin{eqnarray}
\frac{\rho_{oo;ee}}{2}\sin\phi_{oo;ee}=\frac{t_xt_y}{2}\biggl\{
-\sin(\varphi_x-\varphi_y)+
\exp\biggl(\frac{-\Delta^2}{2w_0^2}\biggr)\biggl(
r_x\sin(3\varphi_x-\varphi_y)-
r_y\sin(3\varphi_y-\varphi_x)\biggr)+\nonumber\\
r_xr_y\biggl[-\sin(3[\varphi_x-\varphi_y])+
\exp\biggl( \frac{-\Delta^2}{2w_0^2}\biggr)
\biggl(r_y\sin(5\varphi_y-3\varphi_x)-
r_x\sin(5\varphi_x-3\varphi_y)\biggr)+
\nonumber\\
\exp\biggl(\frac{-2\Delta^2}{w_0^2}\biggr)
\biggl(\frac{r_y}{r_x}\sin(5\varphi_y-\varphi_x)-
\frac{r_x}{r_y}\sin(5\varphi_x-\varphi_y)\biggr)+\nonumber\\
\exp\biggl( \frac{-9\Delta^2}{2w_0^2}\biggr)
\biggl(\frac{r_y^2}{r_x}\sin(7\varphi_y-\varphi_x)-
\frac{r_x^2}{r_y}\sin(7\varphi_x-\varphi_y)\biggr)
\biggr]+\cdots
\biggr\}\nonumber
\end{eqnarray}
which leads to Eq. (A16c) of Ref. \citeonline{bretagne}.

Let us mention that, as stated in section \ref{sec-scalar},  
 Eq. (\ref{rennesm}) represents the Jones matrix of a quarter wave plate
 only under the approximation $\rho_{oo,ee}\approx\rho_{oo}\rho_{ee}$.

\section{Appendix II}

In this appendix, ingredients for the calculation of the double integral
 of Eq. (\ref{master}) are given. 

Following Ref. \citeonline{yeh2x2-ax} we write 
\[
\chat=\sin\theta_c\cos\phi_c\xhatI
+\sin\theta_c\sin\phi_c\yhatI+\cos\theta_c\zhatI
\]
for the direction of the optical axis inside the Quartz plate.
 $\{\xhatI,\yhatI,\zhatI\}$ is the basis attached to the Quartz plate (see 
 Fig. \ref{zigzag}). In this basis, the wave vectors of the plane wave
 (see Eq. (\ref{kperp})) and Gaussian beam center are given by:
\begin{align}
{\rm\bf k_{pw}}&=k_x\xhatI+(k_y\cos\theta_1+k_z\sin\theta_1)\yhatI
+(k_z\cos\theta_1-k_y\sin\theta_1)\zhatI\nonumber\\
{\rm\bf k}&=k\sin\theta_1\yhatI+k\cos\theta_1\zhatI\nonumber
\end{align}
with $k_z\approx k(1-\kperpd/(2k^2)))$.
 The plane wave incident and azimuth angles read
\[
\cos\theta_{1pw}=\frac{k_z}{k}\cos\theta_1-
\frac{k_y}{k}\sin\theta_1,
\,\,
\tan\phi_{pw}=\frac{k_z\sin\theta_1+k_y\cos\theta_1}{k_x}.
\]


The ordinary and extraordinary wave vectors corresponding to 
${\rm\bf k_{pw}}$ and $\chat$
 are determined using the compact expression of Ref. \citeonline{yeh2x2-ax}.
 The electric polarization vectors inside the plate are determined using the 
 general formula of Ref. \citeonline{yeh4x4}.

Expressions for the
 interface matrices of Eqs. (\ref{coeff-1}-\ref{coeff-7}) are formally 
 determined thanks to the Mapple 
 software package \cite{mapple}. These expressions are much too lengthy
 to be reproduced here. 

The rotation matrix $\Omega$ (see Eq. (\ref{omega})) is obtained from 
\begin{eqnarray}
\shatpw&=&N_{s}[(k_y\cos\theta_1+k_z\sin\theta_1)\xhat
-k_x\cos\theta_1\yhat-k_x\sin\theta_1\zhat]\nonumber\\
\phatpw&=&N_{p}[(k_xk_z\cos\theta_1-k_xk_y\sin\theta_1)\xhat
+(k_yk_z\cos\theta_1+k_z^2\sin\theta_1+k_x^2\sin\theta_1)\yhat\nonumber\\
& &+(-\kperp^2\cos\theta_1-k_yk_z\sin\theta_1)\zhat]\nonumber\\
\khatpw&=&N_{k}[k_x\xhat+k_y\yhat+k_z\zhat]\nonumber
\end{eqnarray} 
where $N_s$, $N_p$ and $N_k$ are normalization factors.


\newpage

\begin{thebibliography}{9}

\bibitem{azzam} R.M.A. Azzam and N.M. Bashara, 
{ Ellipsometry and polarized light} (Amsterdam, North-Holland, 1977).

\bibitem{bretagne} J. Poirson et al., ``Jones matrix of a quarter-wave plate 
 for Gaussian beams'', Appl. Opt. 34, 6806-6818 (1995).

\bibitem{fourier-scal} E.C.G. Sudarshan, R. Simon and N. Mukunda, 
 ``Paraxial-wave optics and relativistic front description.
 I. The scalar theory'' Phys. Rev. A 28, 2921-2932 (1983).


\bibitem{fourier-vec} N. Mukunda, R. Simon and E.C.G. Sudarshan,   
 ``Paraxial-wave optics and relativistic front description.
 I. The vector theory'', Phys. Rev. A 28, 2933-2942 (1983).


\bibitem{yeh4x4} P. Yeh, ``Electromagnetic propagation in birefringent media'',
 J. Opt. Soc. Am. 69, 742-756 (1979);  P. Yeh, ``Optics of anisotropic layered
 media: a new $4\times 4$ matrix algebra'', Surf. Sci. 96, 41-53 (1980).



\bibitem{cross} R. Simon, E.C.G. Sudarshan and N. Mukunda, ``Cross polarization
 in laser beams'', Appl. Opt. 26, 1589-1593 (1987).

\bibitem{bacry} H. Bacry and M. Cadihac, ``Metaplectic group and Fourier
 optics'', Phys. Rev. A 23, 2533-2536 (1981).  

\bibitem{simon} Y. Fainman and J. Shamir, ``Polarization of nonplanar wave
 fronts'', Appl. Opt. 23, 3188-3195 (1984).

\bibitem{linear} N. Mukunda, R. Simon and E.C.G. Sudarshan,   
 ``Paraxial Maxwell beams: transformation by general linear optical systems'',
 J. Opt. Soc. Am. A 2, 1291-1296 (1985).

\bibitem{huard} S. Huard, { Polarisation de la lumi\`ere} 
(Masson, Paris, 1993).

\bibitem{handbook} M. Bass et al. {\it Handbook of optics, Vol. II}
 (Mc Graw-Hill, New-York, 1995).


\bibitem{zhu} X. Zhu, ``Explicit Jones transformation matrix
 for a tilted birefringent plate with its optic axis parallel to the plate
 interface'', Appl. opt. 33, 3502-3506 (1994).

\bibitem{yeh2x2}  P. Yeh, ``Extended Jones matrix method'',
 J. Opt. Soc. Am. 72, 507-513 (1982).

\bibitem{yeh2x2-ax} C. Gu and P. Yeh, ``Extended Jones matrix method. II'',
  J. Opt. Soc. Am. A10, 966-973 (1993).

\bibitem{zander} K. Zander, J. Moser and H. Melle, ``Change of polarization
 of linearly polarized, coherent light transmitted through plane-parallel
 anisotropic plates'', Optik 70, 6-13 (1985).

\bibitem{mapple} Maple V software program (Waterloo Maple Inc.,
 Ontario, Canada).  


\end{thebibliography}
\end{document}